\documentclass{PoS}

\title{CMS pixel upgrade project}

\ShortTitle{CMS pixel upgrade project}

\author{\speaker{Hans-Christian K\"astli}\thanks{On behalf of the CMS pixel community.}\\
        Paul Scherrer Institut\\
        E-mail: \email{hans-christian.kaestli@psi.ch}}

\abstract{The LHC machine at CERN finished its first year of pp collisions at a center of mass energy of 7~TeV. While the commissioning to exploit its full potential is still ongoing, there are plans to upgrade its components to reach instantaneous luminosities beyond the initial design value after 2016. A corresponding upgrade of the innermost part of the CMS detector, the pixel detector, is needed. A full replacement of the pixel detector is planned in 2016. It will not only address limitations of the present system at higher data rates, but will aggressively lower the amount of material inside the fiducial tracking volume which will lead to better tracking and b-tagging performance. This article gives an overview of the project and illuminates the motivations and expected improvements in the detector performance.
}

\FullConference{19th International Workshop on Vertex Detectors - VERTEX 2010\\
		June 06 - 11, 2010\\
		Loch Lomond, Scotland, UK}

\begin{document}

After the technical commissioning in 2009, the CMS detector \cite{ref:cms} just finished its first year of physics data taking with good success. Its tracking device consisting of a pixel vertex detector and roughly 200 m$^2$ of silicon strip detectors performs as expected and in many respects reached already the design values \cite{ref:performance}. \\
The commissioning of the LHC machine progresses very well and a peak luminosity at the design value of $10^{34}$ cm$^{-2}$ s$^{-1}$ is expected before 2015. In the following shut down the LHC machine and its injector chain will be upgraded in order to reach at least twice this value. This is called the phase I upgrade. A phase II upgrade is under consideration. Its time scale is much more uncertain but it is expected to take place in the early 2020s where the LHC should reach a luminosity larger than $5\times10^{34}$ cm$^{-2}$ s$^{-1}$.\\
Corresponding upgrades of the CMS detector are necessary. This article gives an overview of the phase I upgrade of the CMS pixel detector. The phase II upgrade of the entire tracker is still in its early conceptual phase and is not described here.

\section{Motivation}
Once the LHC machine reaches a mode of operation which goes beyond its initial design goal the CMS tracking detector faces two main issues. 
\begin{enumerate}
    \item The hit occupancy increases. This causes an increase in data loss due to buffer size limitations and limited readout data bandwidth. At an instantaneous luminosity of $2\times 10^{34}$ cm$^{-2}$ s$^{-1}$ this becomes inadequate for the present pixel detector. Changes in the front end electronics and in the data links are needed.
    \item Parts of the pixel detector will have to be replaced before the end of the phase I running period due to radiation induced damage of sensors and readout electronics \cite{ref:trackerTDR}. At the time of writing, the planned total LHC integrated luminosity up to 2020 is $\approx$340~fb$^{-1}$ corresponding to $1.7\times10^{15}$ 1~MeV neutron equivalent for layer 1. In the TDR the modules are specified up to  $6\times10^{14}$ 1~MeV n.e. While the detector will remain efficient for at least twice this dose the spacial resolution will gradually decrease due to the reduction of the Lorentz angle with higher bias voltages \cite{ref:Dorokhov}. 
    \item With higher track densities tracking becomes more and more time consuming. The track fake rate grows rapidly due to the large extrapolation distance from pixel track seeds to the first strip detector layer (see below, i.e. figure \ref{fig:TrackEff}) . In the barrel this is from 10.3~cm (outermost pixel layer) to 25.5~cm (innermost TIB layer). An intermediate pixel layer is highly desirable.
\end{enumerate}

\begin{figure}[ht]
\begin{centering}
 \includegraphics[width=0.65\textwidth]{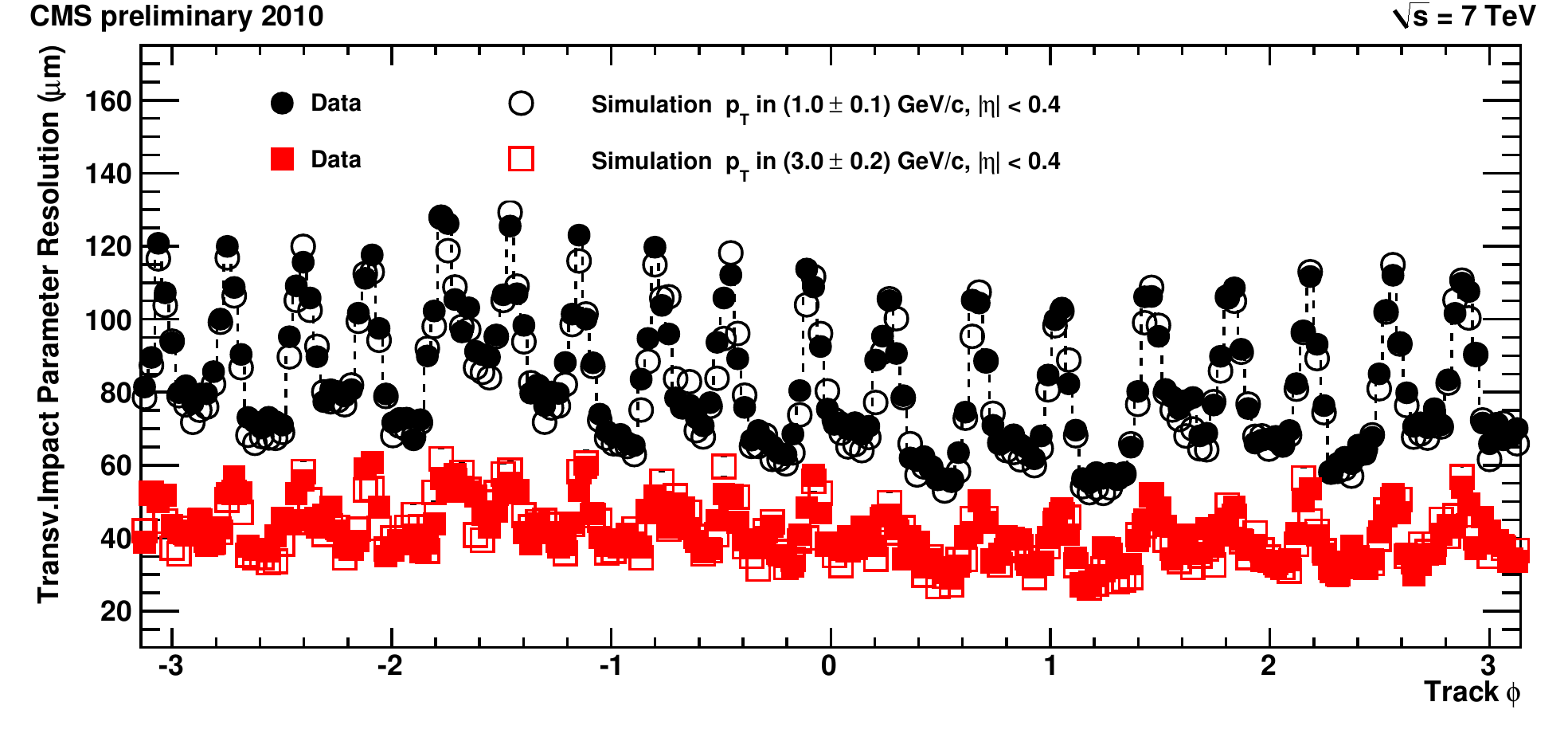}
    \caption{Transverse impact parameter resolution as a function of azimuthal angle $\phi$ of the track measured with the current CMS detector. At low transverse momenta the material effect of the cooling pipes are clearly visible as it degrades the resolution due to multiple scattering.}
    \label{fig:ip_res_old}
\end{centering}
\end{figure}

In addition there are physics driven arguments for an upgrade. 
\begin{enumerate}
    \item[4.] Due to multiple scattering any material inside the fiducial tracking volume decreases the impact parameter resolution. This is illustrated in figure \ref{fig:ip_res_old}. For low momentum tracks (shown are tracks around 1 and 3 GeV) 18 peaks are clearly visible in the transverse impact parameter resolution as a function of $\phi$, which correspond to the 18 cooling pipes of the innermost pixel barrel layer. The upgraded pixel detector features a mechanical structure which is consequently tuned for low mass.
    \item[5.] For purely geometrical reasons a reduction in the radius of the innermost tracking layer will further improve the impact parameter resolution. A reduction from today's 44~mm to 39~mm is foreseen and further reductions with smaller beam pipe radii are currently under study. 
\end{enumerate}

\section{Mechanical design}

\begin{figure}[hb]
\begin{centering}
 \includegraphics[width=0.99\textwidth]{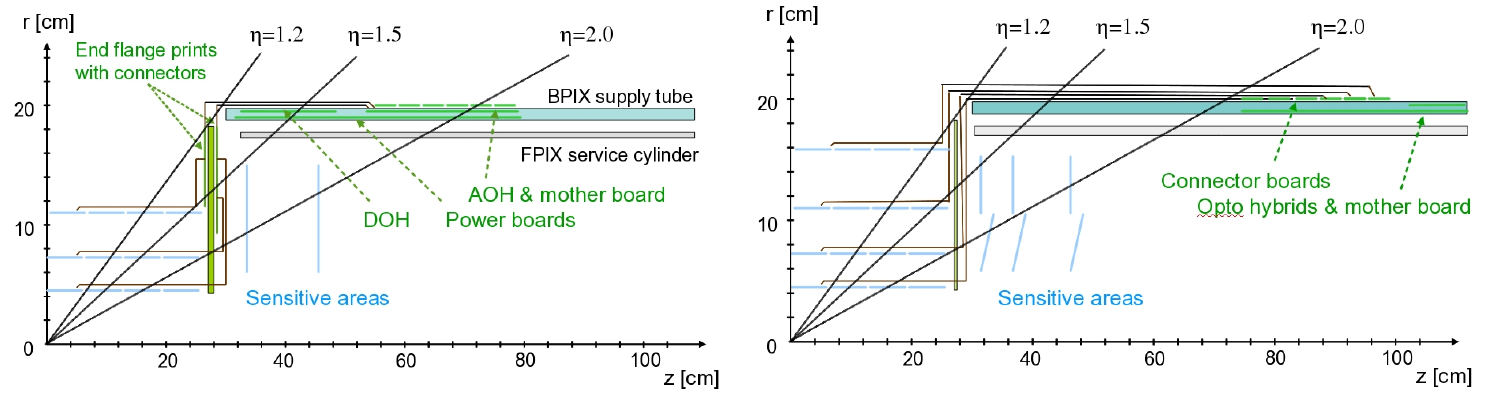}
    \caption{Side view of the present (left) and future (right) pixel system. Emphasis has been given to move as much material as possible out of the fiducial tracking region and hence to higher $\eta$ values. The amount of material in the region $|\eta |<2.16$ will be reduced by a factor 2.6 for the barrel pixel.}
    \label{fig:outline}
\end{centering}
\end{figure}

The proposed upgraded pixel system is sketched in figure \ref{fig:outline} right. It consists of three disks on each side and four barrel layers. For comparison the present 2+2 disk/3 barrel layer system is shown on the left. The large reduction in material can be achieved based on 2 main ideas:
\begin{itemize}
    \item Change of the cooling system to a 2 phase CO$_2$ cooling. The present C$_6$F$_{14}$ monophase cooling with its piping accounts for approximately $1/3$ of the total material budget per layer in the central region \cite{ref:stefan}. The benefit of CO$_2$ is twofold. First, the mass density of the bi-phase CO$_2$  is lower than for C$_6$F$_{14}$. Second, with the high latent heat of the CO$_2$ much less mass flow at a higher pressure is needed leading to smaller pipes. Pipes with a diameter of 1.5~mm and a wall thickness of 50~$\mu$m are foreseen. The amount of material needed for cooling (pipes plus coolant) for the ladders of the entire barrel pixel detector will go down by a factor of 10 to about 164~g.
    \item Relocation of material out of the fiducial tracking volume towards higher $|\eta|$ values and complete removal of circuit and connector boards from the barrel end flange region. Barrel modules will have longer pigtail cables which reach out to a region of the supply tube above $|\eta|=2.1$. 
\end{itemize}

\begin{figure}[htb]
\begin{centering}
 \includegraphics[width=0.44\textwidth]{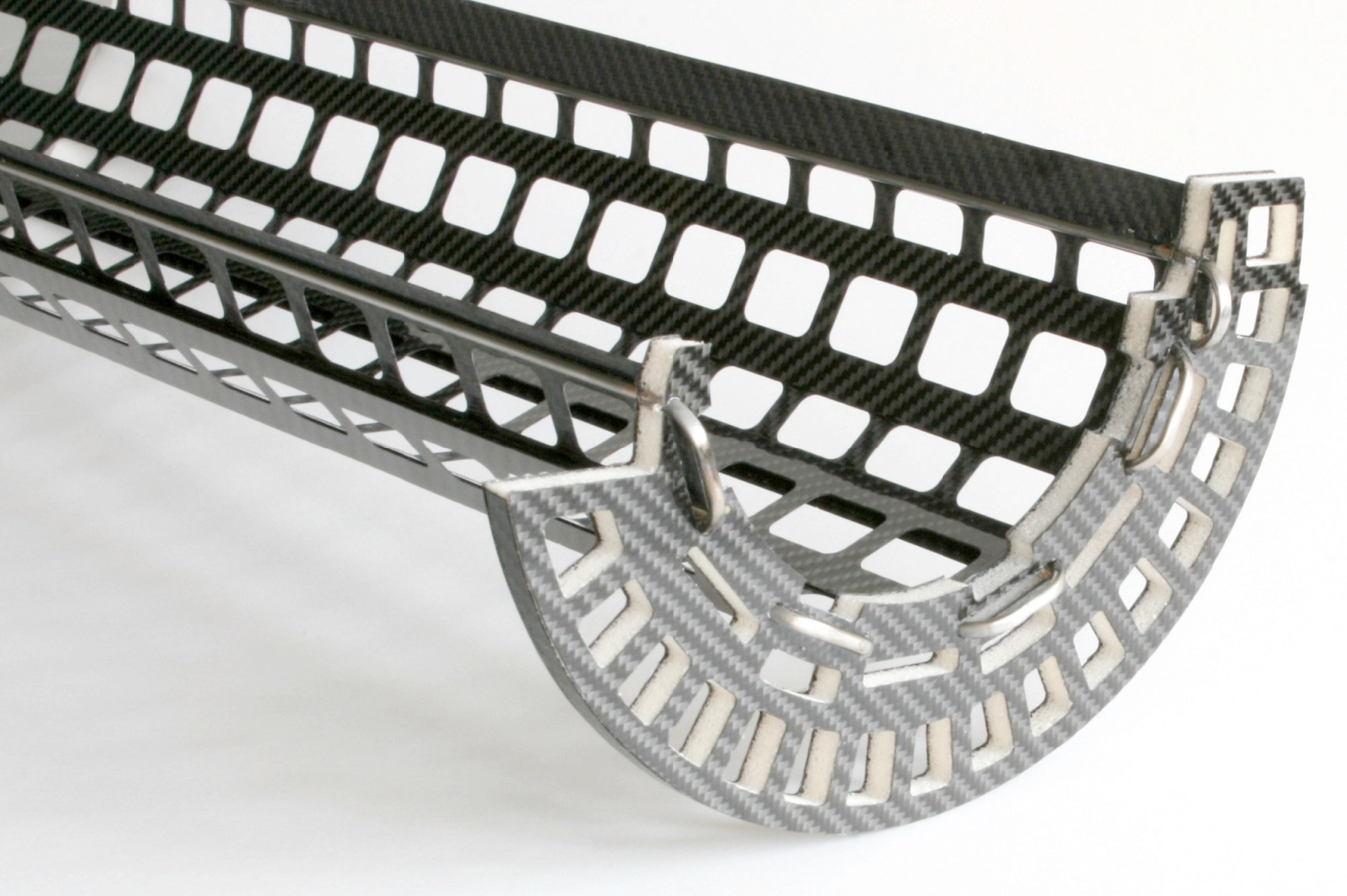}
 \includegraphics[width=0.55\textwidth]{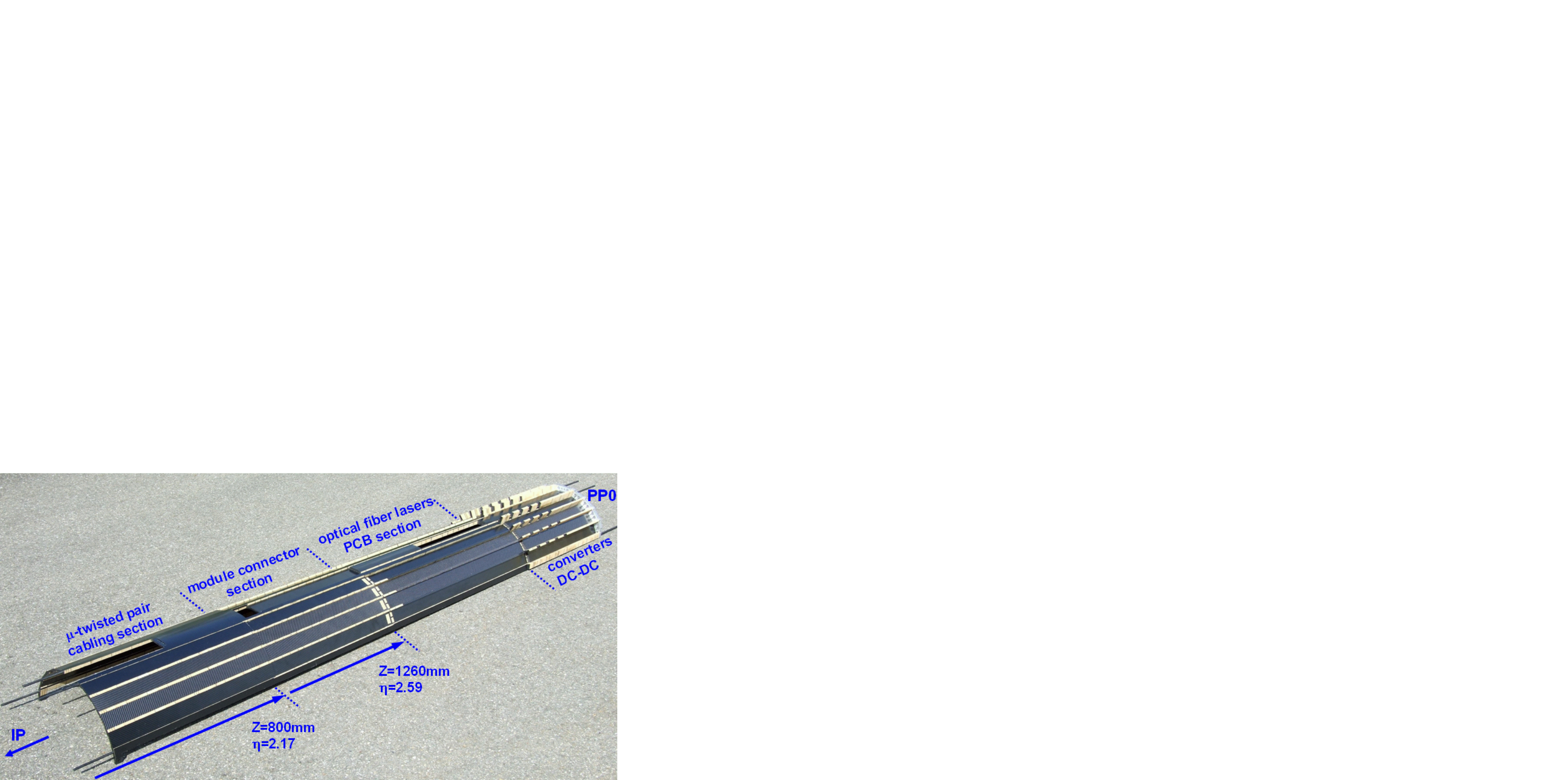}
    \caption{Photographs of prototypes of the first barrel layer mechanics (left) and barrel supply tube (right). It has consequently been designed for low mass. The materials used are carbon fibre sheets glued onto stainless steel tubes for the detector mechanics and carbon fibre and AIREX foam profiles glued together and laminated with carbon fibre sheets for the supply tube.}
    \label{fig:Supplytube}
\end{centering}
\end{figure}

\begin{figure}[htb]
\begin{centering}
 \includegraphics[width=0.49\textwidth]{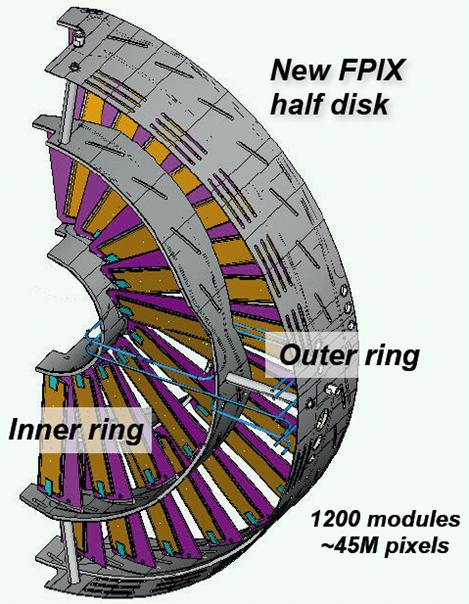}
 \includegraphics[width=0.49\textwidth]{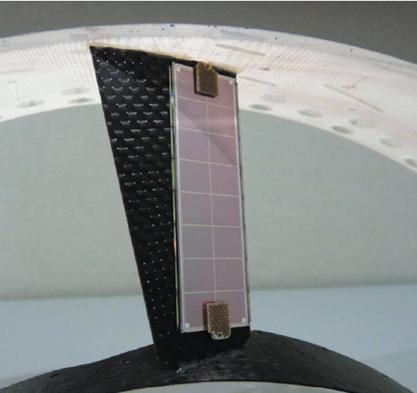}
    \caption{Design of a pixel detector half disk. It consists of an inner and an outer ring (left). The picture on the right shows a photograph of a carbon fibre blade with a module mounted on top.}
    \label{fig:FPIX}
\end{centering}
\end{figure}

A prototype of the first layer barrel mechanics and its supply tube has been built. Photographs of the two objects are shown in figure \ref{fig:Supplytube}. A disk of the endcap detector will consist of two inner and two outer halfrings (see figure \ref{fig:FPIX}). It will consist of only one type of modules consisting of 16 readout chips. On an outer (inner) half ring 34 (22) modules will be mounted. More details can be found in \cite{ref:hadi}, \cite{ref:stefan}.

\begin{figure}[htb]
\begin{centering}
 \includegraphics[width=0.49\textwidth]{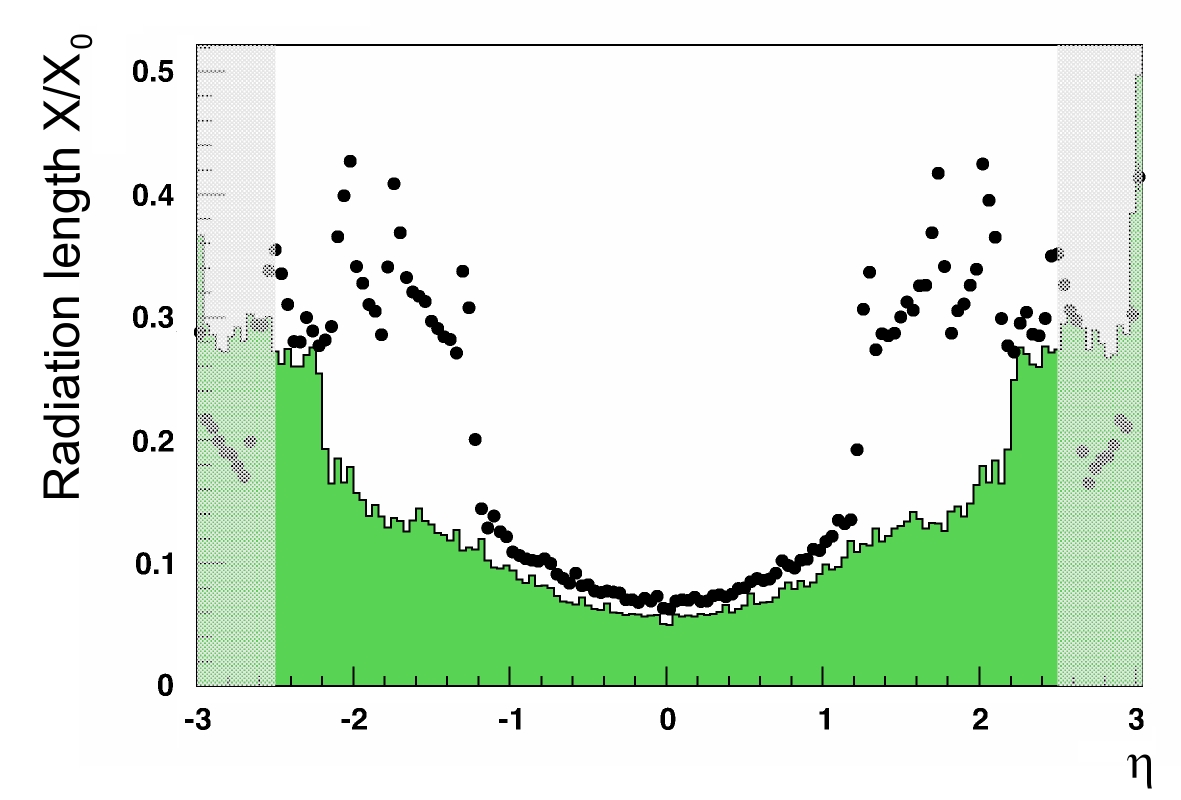}
 \includegraphics[width=0.49\textwidth]{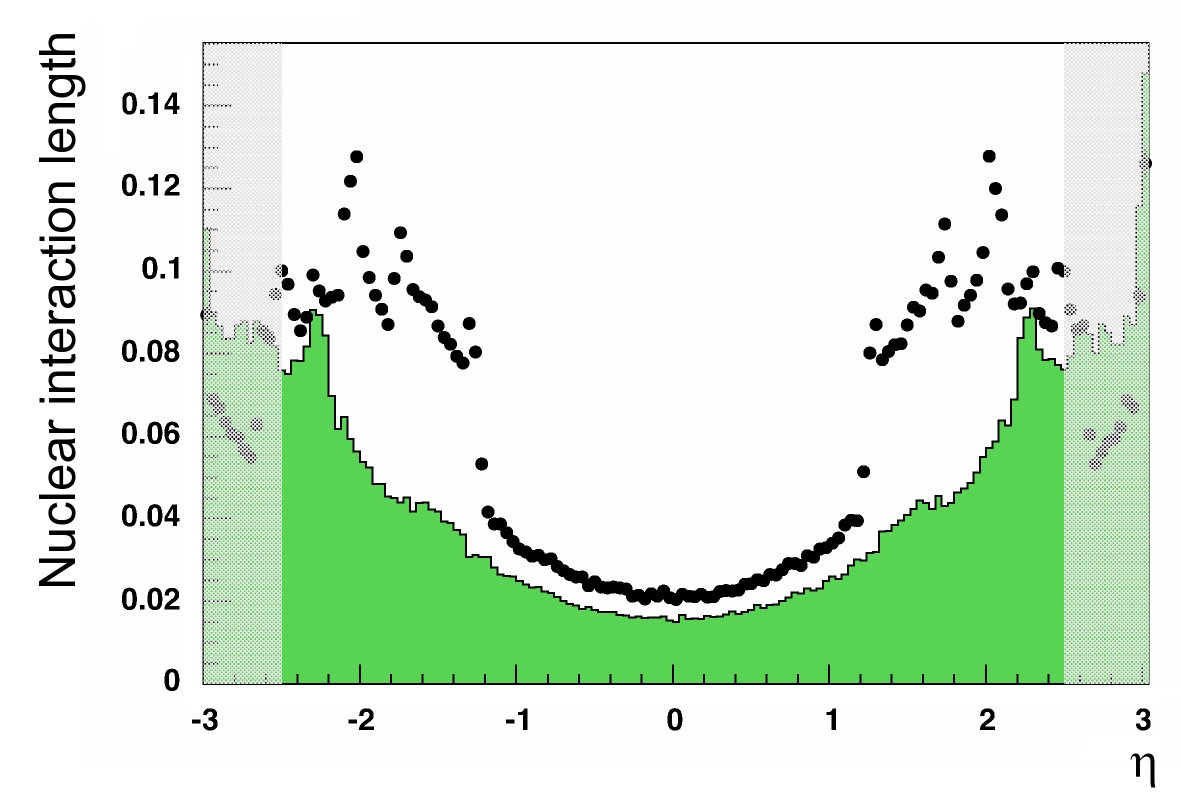}
    \caption{Comparison of the material budget of the existing 3 layer barrel pixel detector (black points) and the proposed 4 layer detector (green histogram). Shown are the radiation length (left) and nuclear interaction lengths (right) as a function of pseudorapidity $\eta$. The shaded regions are outside the fiducial tracking volume.}
    \label{fig:MaterialBudget}
\end{centering}
\end{figure}

The expected distribution of material in the pixel detector is plotted in figure \ref{fig:MaterialBudget}. Shown is the thickness of the barrel detector in radiation length (left) and nuclear interaction length (right). In spite of the additional barrel layer, the material budget is reduced by 20~\% at $\eta$=0 and by more than 50~\% in the region $1.4<\vert\eta\vert<2.1$. The amount of material in the region $|\eta |<2.16$ will be reduced by a factor of 2.6 from today's 16.9~kg to 6.5~kg.

\section{Sensor and electronics}

The pixel size of the new design remains at $100\times 150~\mu$m (in r$\phi$-z). The sensor technology will again be ``n-in-n'' planar silicon. The exact choice of sensor material (DOFZ, MCZ, ...) can be postponed to the moment of ordering. \\
The new design has several implications on the readout electronics. It can be divided in two sections: services and readout chips

\subsection{Services}
The number of channels will be increased by almost a factor 2 to 125M pixels (today 66M pixels). While there have been services foreseen for the $3^{\mathrm rd}$ disks from the beginning, the new 4 layer barrel detector has to reuse the same services as the present 3 layer system with the addition of a small number of spares, i.e. the same number of power cables, cooling pipes and optical fibres. This implies the following changes:
\begin{itemize}
    \item Change of cooling system as described above. The present C$_6$F$_{14}$ cooling system will not be able to supply enough cooling capacity to the larger detector system. 
    \item Faster data links. Since the data of twice as many channels need to be transmitted over almost the same number of optical fibres, the link bandwidth needs to be increased. This seems very difficult on the basis of the present analog links. Therefore low power high speed (320~MHz) digital links are under development \cite{ref:beat}. 
    \item Alternative powering scheme. Since considerably more power has to be provided over almost the same number of power cables, the power loss in the cables become unacceptably high unless the power is supplied at a higher voltage. A DC-to-DC conversion scheme has been chosen and is under development \cite{ref:katja}.
\end{itemize}

\subsection{Readout chip}
\begin{figure}[htb]
\begin{centering}
 \includegraphics[width=0.49\textwidth]{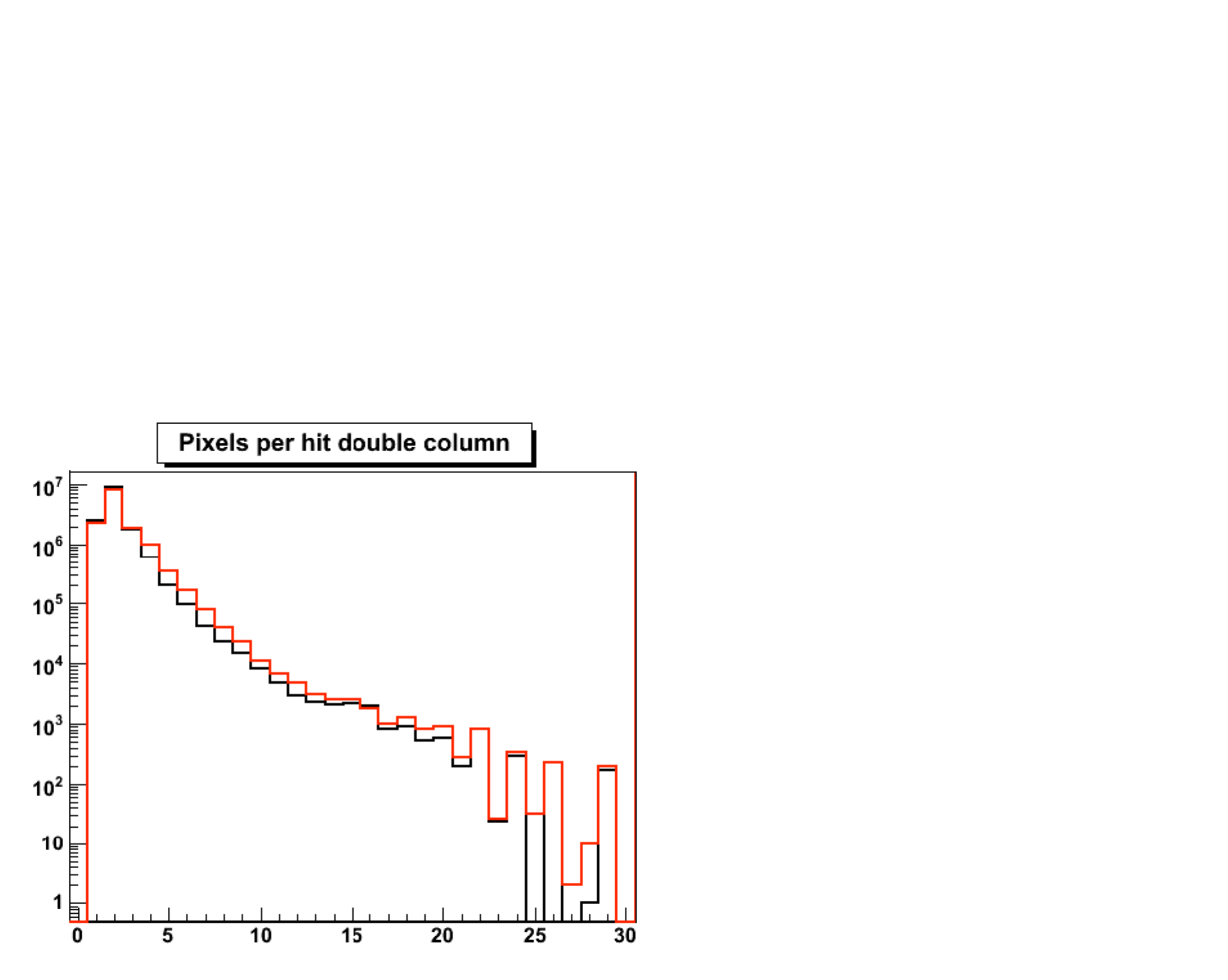}
 \includegraphics[width=0.49\textwidth]{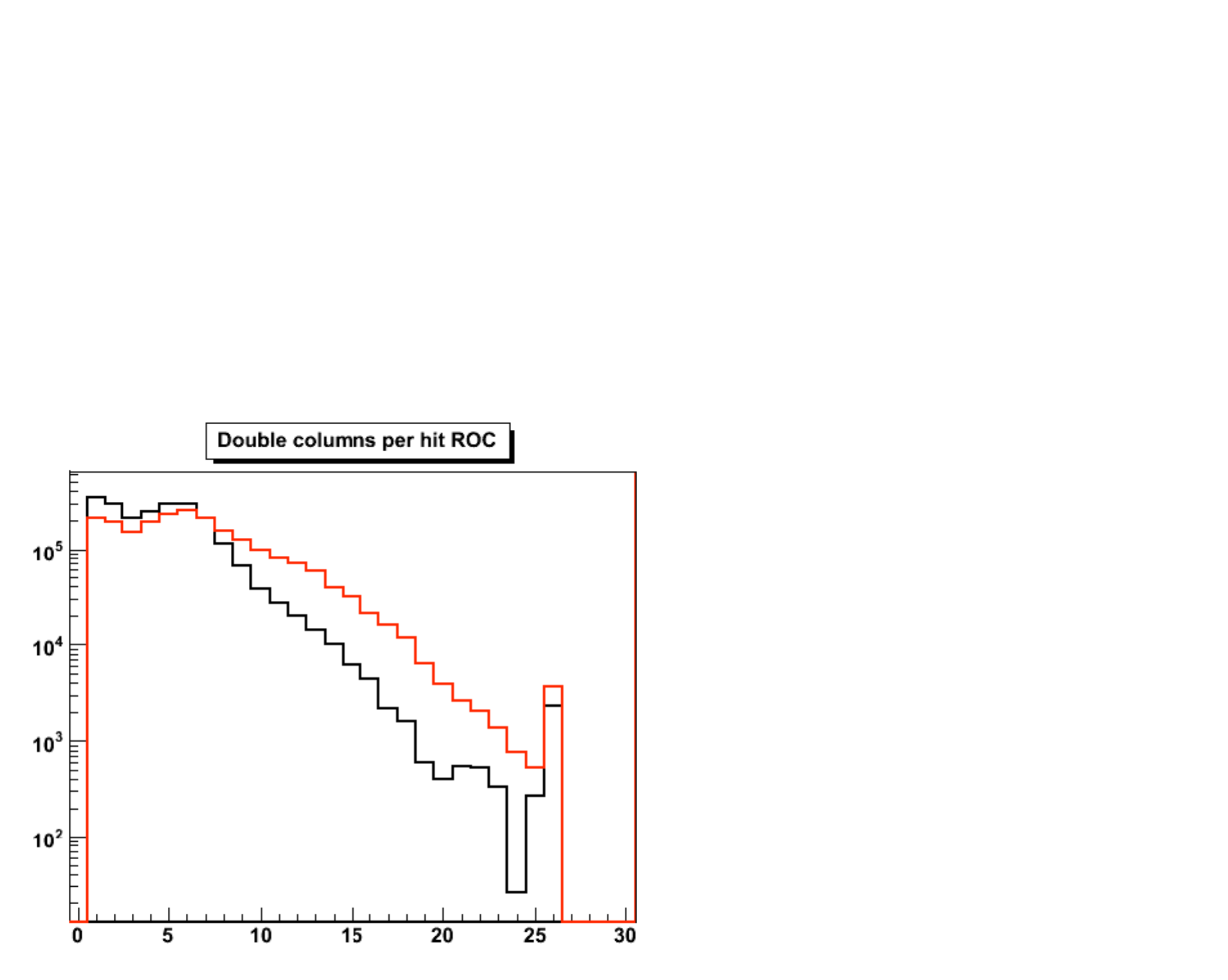}
 \includegraphics[width=0.49\textwidth]{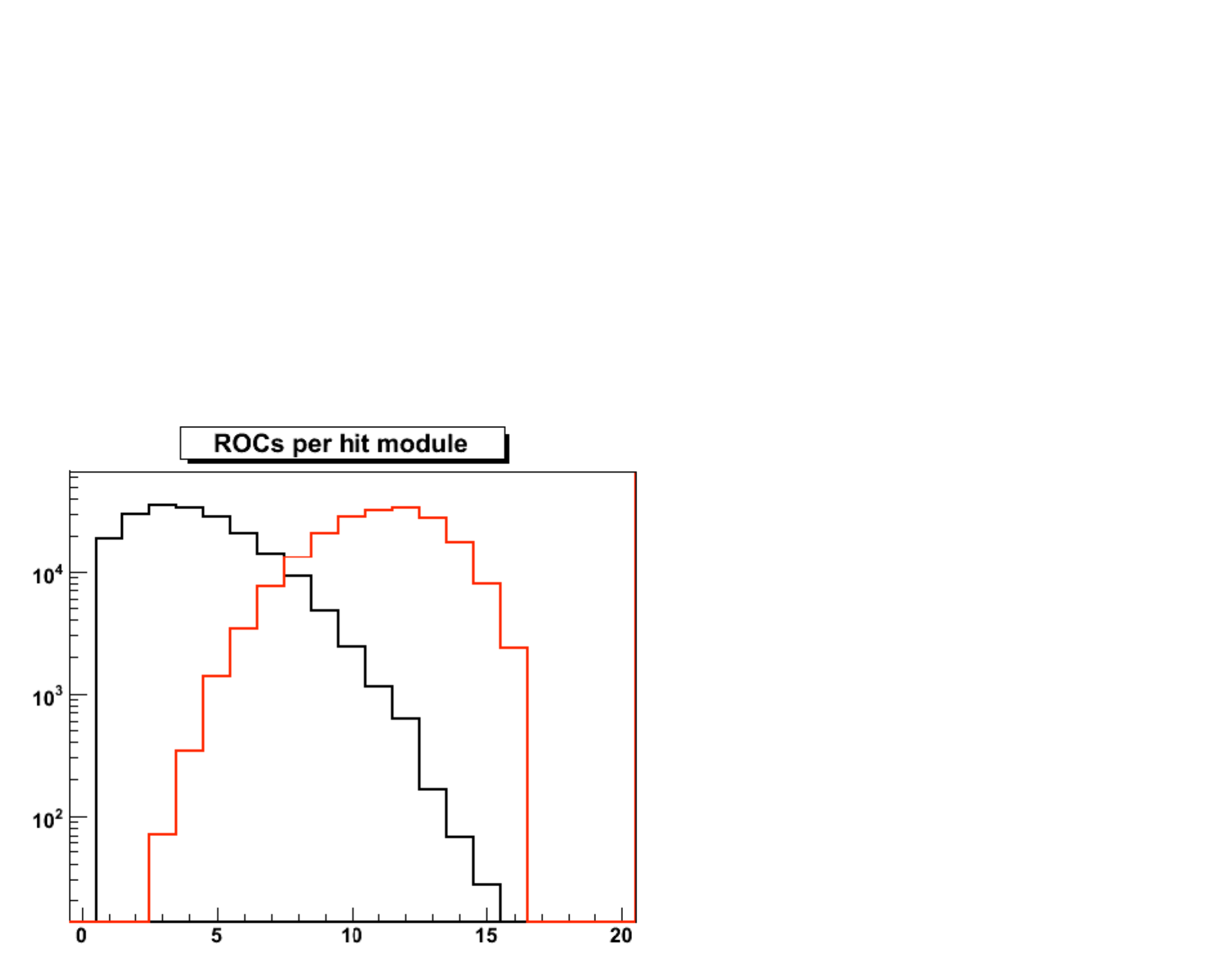}
 \includegraphics[width=0.49\textwidth]{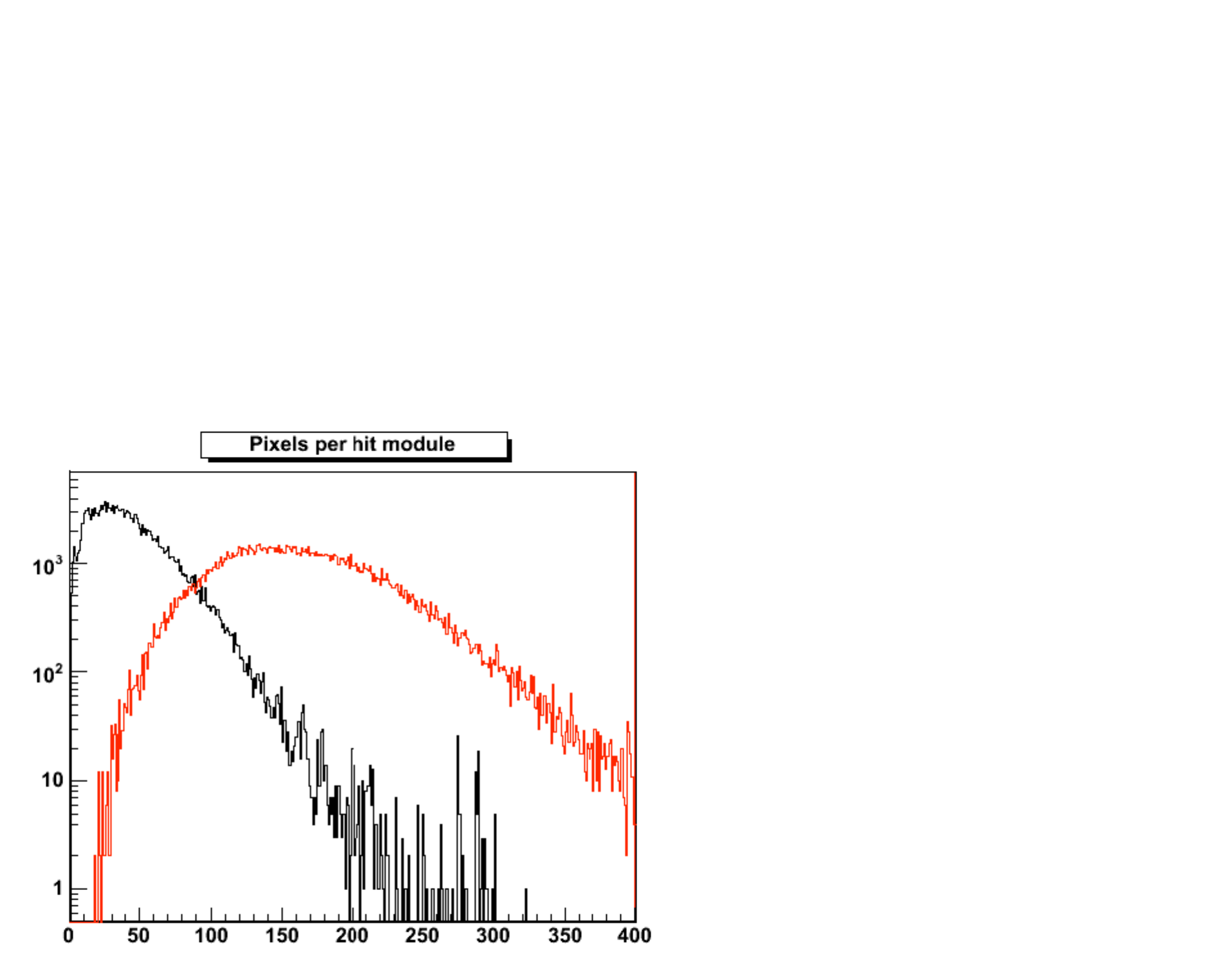}
    \caption{Occupancy in a barrel module for the innermost layer at luminosities of 1 (black) and 2 (red) $\times 10^{34}{\mathrm cm}^{-2}{\mathrm s}^{-1}$. Shown are from top left to bottom right the number of hit pixels per hit double column, the number of hit double columns per hit ROC, the number of hit ROCs per hit module and the number of hit pixels per hit module.}
    \label{fig:Occupancy}
\end{centering}
\end{figure}

The present readout chip (ROC) has been designed to be efficient at the nominal LHC luminosity of $1\times 10^{34}$ cm$^{-2}$ s$^{-1}$ . It is inappropriate for a peak luminosity of $2\times 10^{34}$ cm$^{-2}$ s$^{-1}$. Data flow simulations show, that for the innermost layer the pixel hit recording efficiency would drop below 65~\%. However, the limitations are not inherent to the architecture chosen. Therefore an evolution of the present ROC is under design (\cite{ref:beat}) which will overcome these limitations. There are mainly four aspects:
\begin{enumerate}
    \item {\bf Trigger latency buffers.} Pixel hits need to be stored inside the ROC during the level 1 trigger decision time ($\approx 3~\mu$s). The size of these buffers need to be adjusted to the higher data rates. This has been done.
    \item {\bf Pixel readout.} Trigger validated hits need to be read out. Dead time of a double column (a ROC is organized in 26 double columns with 160 pixels each) starts with a trigger validation and ends when the validated hits are read out. A fast removal of the hits from the double column is therefore desirable. Today, all double columns within a ROC are daisy chained for readout as well as 8 ROCs on a module, aggregating to a readout chain of 208 double columns. Readout time and hence dead time increases along this chain, leading to a mean inefficiency of $\approx 3.8\%$ at the design LHC luminosity. This inefficiency further grows to an unacceptably high value of 16~\% at $2\times 10^{34}$ cm$^{-2}$ s$^{-1}$. The solution is to break up the daisy chain on the module level. The situation is shown in figure \ref{fig:Occupancy}. It shows the length of a readout (in number of pixels) of different structures (double column, ROC, module) for  1 and $2\times 10^{34}$ cm$^{-2}$ s$^{-1}$, the later for the worst case of 50~ns LHC bunch structure or a mean of 100 pile up events per bunch crossing. As expected for uncorrelated pile up events, the local event size (within a ROC) is very similar with a large difference showing up on a larger scale (module level). This justifies to break up the daisy chain between ROCs, leaving the readout scheme inside the ROC unaltered. It requires the addition of a buffer stage on the ROC level. The hits from the double columns are written into this readout buffer instead of sent off detector directly. Hits in this buffer are read out at a later time, not leading to further dead time, since the double column already resumed data taking. From simulations we expect the inefficiency to drop to 6~\% for layer 1 in this new scheme. 
    \item {\bf Digital output.} In order to transmit data faster to the data acquisition units the present analog links are abandoned and changed to 160/320~MHz digital links. Therefore a fast ADC is needed on chip with the corresponding supporting circuits (like a PLL to generate higher clock frequencies or a data serializer and line drivers).
\end{enumerate}

The changes described so far will not alter the core of the ROC which is very well tested and debugged. Notably this includes the pixel cell. This chip is the baseline for the phase I upgrade and a submission is planned for summer 2011. In order to reduce the inefficiency even further changes to the complicated double column logic are needed. It is currently under consideration and eventually will be submitted in 2012.
 
\begin{enumerate}
    \item[4.] {\bf Double column dead time and reset.} In order to eliminate the double column dead time completely a more intelligent logic is needed. Today, the double column stops data taking after trigger validation in order to avoid overwriting of valid data. After readout a reset is issued to ensure synchronisation and data integrity. This could be avoided if the data buffer logic would protect buffer cells containing valid data without stopping data acquisition. However, this means a complete redesign of the double column logic with the corresponding time consuming test phase in laboratory and high rate test beams. It is currently under design at PSI. 
\end{enumerate}

\section{Expected performance}

\begin{figure}[ht]
\begin{centering}
 \includegraphics[width=0.9\textwidth]{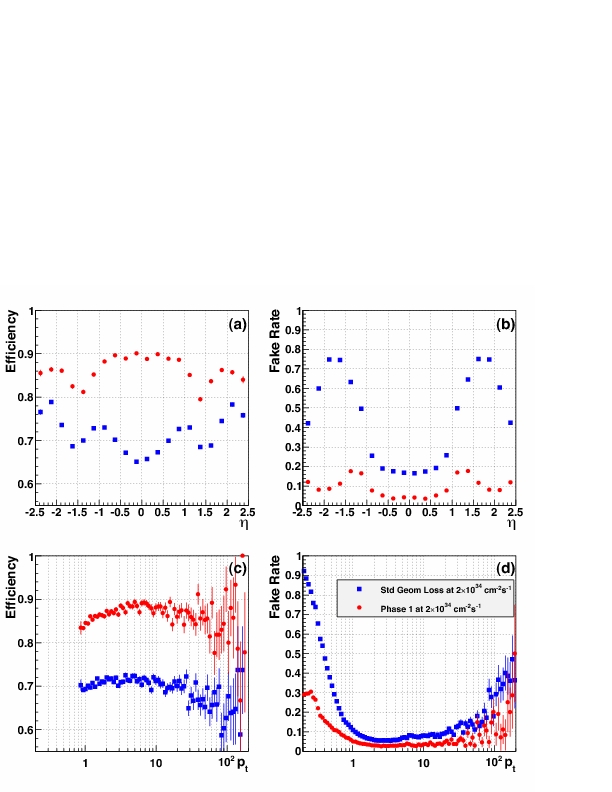}
    \caption{Tracking efficiency and fake rate for the current pixel detector (blue) and the proposed upgraded detector (red) for $t\bar{t}$ events with pile up corresponding to $2\times 10^{34}{\mathrm cm}^{-2}{\mathrm s}^{-1}$ and 25~ns bunch spacing. Shown are: (a) tracking efficiency vs pseudorapidity; (b) fake rate vs pseudorapidity; (c) efficiency vs p$_{\mathrm T}$; (d) fake rate vs p$_{\mathrm T}$. }
    \label{fig:TrackEff}
\end{centering}
\end{figure}

\begin{figure}[ht]
\begin{centering}
 \includegraphics[width=0.6\textwidth]{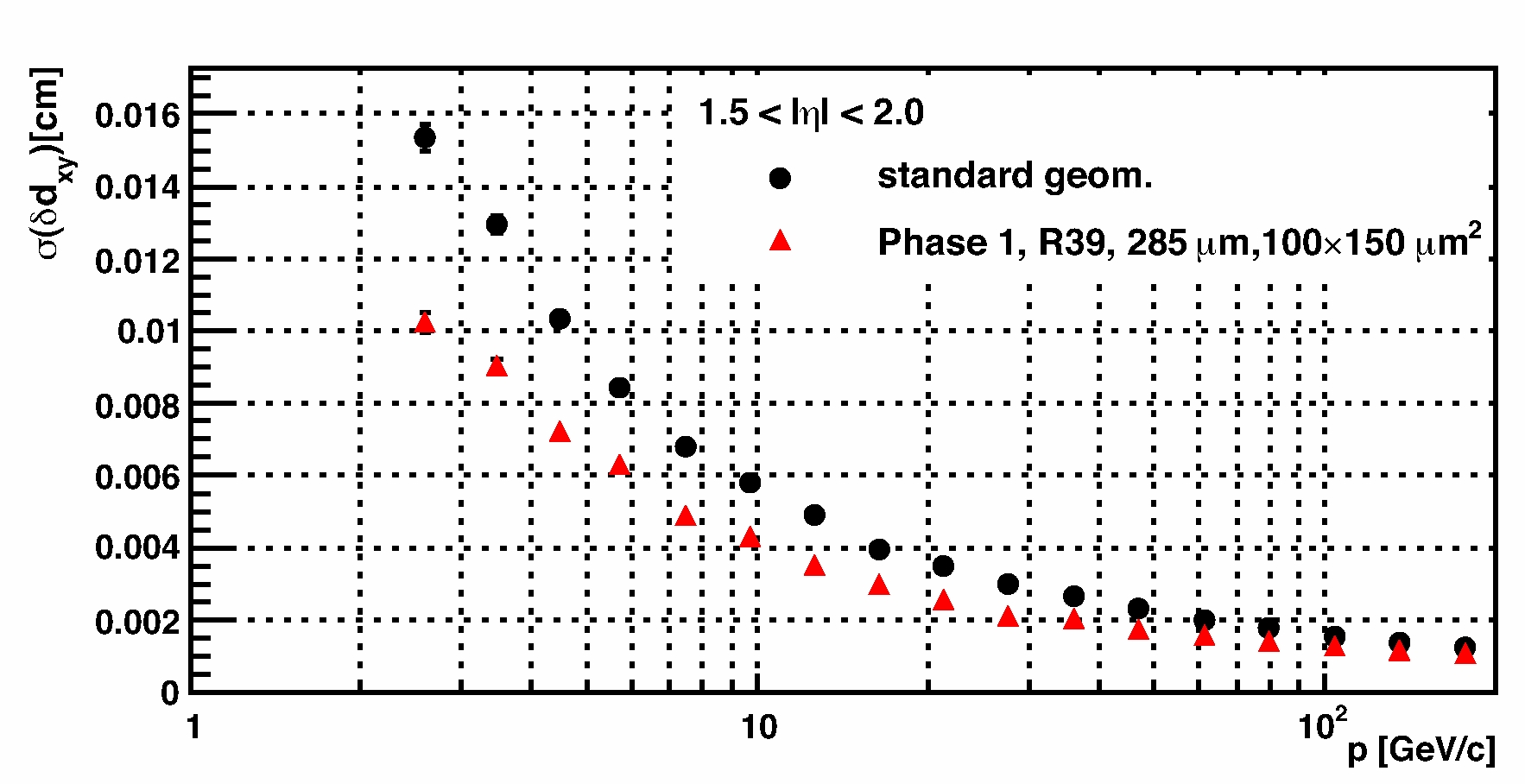}
    \caption{Transverse impact parameter resolution as a function of azimuthal angle $\phi$ of the track measured with the current CMS detector. At low momenta the material effect of the cooling pipes are clearly visible as it degrades the resolution due to multiple scattering.}
    \label{fig:ip_res}
\end{centering}
\end{figure}

The new geometry (as far as it is known today) has been implemented in GEANT4 to simulate the detailed detector response within the official CMS software framework. The tracking efficiency together with the track fake rate as a function of $\eta$ and p$_{\mathrm T}$ is shown in figure \ref{fig:TrackEff} for $t\bar{t}$ events with a number of pile up interactions corresponding to an instantaneous luminosity of $2\times 10^{34}{\mathrm cm}^{-2}{\mathrm s}^{-1}$. As can be seen with the new geometry the tracking efficiency will be improved by about 20\% and at the same time the track fake rate will be drastically reduced by up to an order of magnitude. This is due to the 4 hit coverage up to $\eta\approx2.5$ where pixel hit quadruples and triplets are used for track seeding.\\
The drastic reduction in material within the fiducial tracking region is expected to lead to an improved impact parameter resolution. Figure \ref{fig:ip_res} shows the simulated transverse impact resolution of the old and new system as a function of p$_{\mathrm T}$ in the transition region between barrel and forward pixel detectors. The gain in resolution is of the order of 30\% for tracks with a p$_{\mathrm T}$ around 5~GeV.

\begin{figure}[htb]
\begin{centering}
 \includegraphics[width=0.5\textwidth]{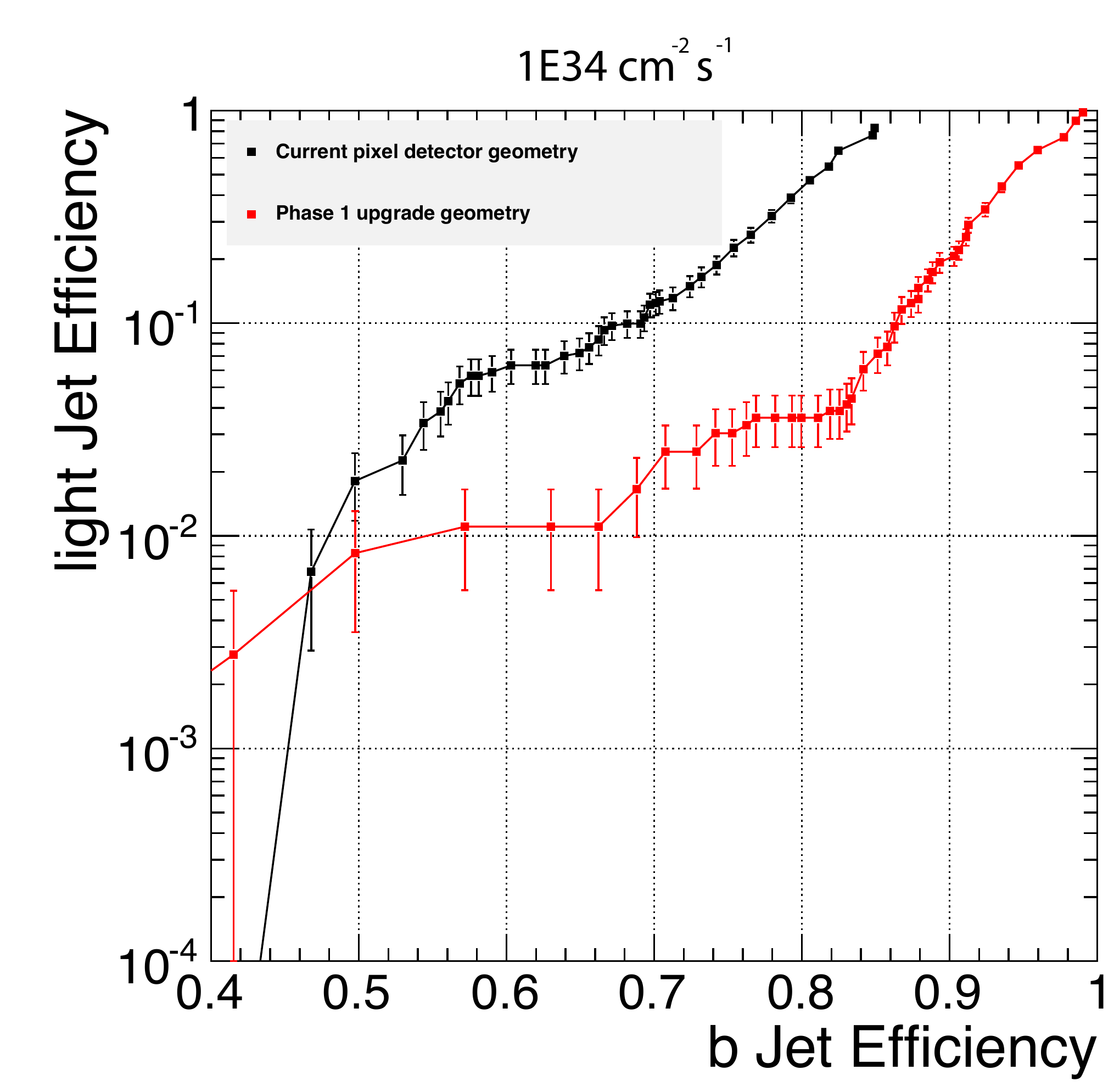}
    \caption{B tagging performance for the old (black) and new (red) pixel detector. For 70~\% b jet tagging efficiency, the rejection factor for light jets increases from $\lesssim 10$ to about 50. }
    \label{fig:b_tag}
\end{centering}
\end{figure}

Finally preliminary studies of the b-tagging efficiency have been carried out. In figure \ref{fig:b_tag} the performance of the standard b-tagging algorithm \cite{ref:b-tag} in CMS is shown. At $70~\%$ b jet tagging efficiency the light jet rejection rate will go up from today 90~\% to about 98~\% for the proposed system. Or conversely at a constant light jet rejection rate of 90~\% the signal efficiency for b jets will increase by 23~\% from 70~\% to 86~\%. 
 
\end{document}